# Newly synthesized MAX phase Zr$_2$SeC: DFT insights into physical properties towards possible applications

M. A. Ali[a,*], Muhammad Waqas Qureshi [b,c]

[a]Department of Physics, Chittagong University of Engineering and Technology (CUET), Chattogram-4349, Bangladesh
[b] State Key Laboratory of Advanced Welding and Joining, Harbin Institute of Technology, Harbin 150001, China
[c]School of Materials Science & Engineering, Harbin Institute of Technology, Harbin 150001, China

**ABSTRACT**

A DFT study of the synthesized MAX phase Zr$_2$SeC has been carried out for the first time to explore its physical properties for possible applications in many sectors. The studied properties are compared with prior known MAX phase Zr$_2$SC. The structural parameters (lattice constants, volume, and atomic positions) are observed to be consistent with earlier results. The band structure and density of states (DOS) are used to explore the metallic conductivity, anisotropic electrical conductivity, and the dominant role of Zr-$d$ states to the electrical conductivity. Analysis of the peaks in the DOS and charge density mapping (CDM) of Zr$_2$SeC and Zr$_2$SC revealed the possible variation of the mechanical properties and hardness among them. The mechanical stability has been checked using elastic constants. The values of the elastic constants, elastic moduli, and hardness parameters of Zr$_2$SeC are found to be lowered than those of Zr$_2$SC. The anisotropic behavior of the mechanical properties has been studied and analyzed. Technologically important thermodynamic properties such as the thermal expansion coefficient(*TEC*), Debye temperature ($\theta_D$), entropy (S), heat capacity at constant volume ($C_v$), Grüneisen parameter ($\gamma$) along with volume (V) and Gibbs free energy (G) are investigated as a function of both temperature (from 0 to 1600 K) and pressure (from 0 to 50 GPa). Besides, the $\Theta_D$, minimum thermal conductivity ($K_{min}$), melting point ($T_m$), and $\gamma$ have also been calculated at room temperature and found to be lowered for Zr$_2$SeC compared to Zr$_2$SC owing to their close relationship with the mechanical parameters. The value of the $\Theta_D$, $K_{min}$, $T_m$, and *TEC* suggest Zr$_2$SeC as a thermal barrier coating material. The optical properties such as dielectric constant (real and imaginary part), refractive index, extinction coefficient, absorption coefficient, photoconductivity, reflectivity, and loss function of Zr$_2$SeC are computed and analyzed to reveal its possible applications.

## 1. Introduction

Since the 1990s, one of the widely known classes of transition metal carbides or nitrides is the so-called MAX phase with general formula M$_{n+1}$AX$_n$ (M – early transition metal, A – A-group

Email for correspondence: ashrafphy31@cuet.ac.bd

element, X – C or N; n = 1- 4) [1–4]. The MAX phase materials are able to attract great attention owing to their exceptional performance combining both ceramics (elastically rigid, lightweight, creep and fatigue resistant as ceramic materials) and metals (machinable, electrically and thermally conductive, not susceptible to thermal shock, plastic at high temperature and exceptionally damage-tolerant) [1,2]; consequently, they are promising candidates such as in high-temperature technology as components, sliding electrical contacts, and contacts for 2D electronic circuits, Li-ion batteries, wear and corrosion-resistant coatings, superconducting materials, spintronics, and nuclear industry.[2,5–10]The hybrid properties of MAX phases are due to strong covalent M–X bonds and relatively weak metallic M–A bonds within their structure[2,11,12]. These challenging properties are always motivating scientists; consequently, more than 150 MAX phases have been discovered. [1]. Moreover, researchers are also trying to manipulate the composition and structure to achieve better combination of the properties such as different alloys/solid solutions [13–21], $M_2A_2X$ [22–24] and $M_3A_2X_2$[25], rare-earth i-MAX phases [26,27], 212 MAX phases [28–30], 314 MAX phases [28,31], MAX phase borides [6,32–38]and two-dimensional (2D) MAX phase derivatives termed MXenes [39,40].

As an A element, chalcogen S containing MAX phases have drawn attention owing to their comparatively strong M-A(*p-d*) bonding which is usually weak for MAX phases. So far known, the S-containing ternary MAX phases are $M_2SC$ (M = Ti, Zr, Hf, and Nb) and $M_2SB$ (M = Zr, Hf, and Nb) [1, 13-15]. The existence of strong M-A bonding for the S-containing MAX phases is reflected from the higher values of Young's modulus, bulk modulus, and shear modulus of $M_2SC$ (M = Ti, Zr, Hf) compared to that of $M_2AlC$ (M = Ti, Zr, Hf)[35,41–43].All the S-containing phases have been studied completely by the different research groups. Amini et al.[41] have synthesized and studied the mechanical properties of $Ti_2SC$. Bouhemadou et al.[42] have performed a first-principles investigation of $M_2SC$ (M=Ti, Zr, Hf) compounds. A comprehensive study of elastic properties of 211 MAX phases has been carried out by Cover et al.[43], where the $Ti_2SC$ and $Zr_2SC$ were included. $M_2SB$ borides have been synthesized by Rackl et al.[33,34], which has been further subjected for the comprehensive study of the physical properties of $M_2SX$ (X = C and B) [35]. For each case, the S-containing MAX phases have enhanced mechanical properties compared to corresponding Al-containing MAX phases.

Recently, a new chalcogen (Se) containing MAX phase ($Zr_2SeC$) has been synthesized by Chen et al. [44], where they have investigated the only electronic density of states, charge density, electrical resistivity, and thermal conductivity of $Zr_2SeC$. These limited results are not enough to explore the $Zr_2SeC$ thoroughly for practical application in many sectors where many other MAX phases have already been used. For example, the MAX phases are using in many sectors as structural components where the information of mechanical properties is essential. MAX phases are also using in high-temperature technology where some prior knowledge of Debye temperature, minimum thermal conductivity as well as melting temperature is required. Moreover, the MAX phases are potential candidates for use as cover materials to diminish solar heating in many sectors where the knowledge of optical properties is fundamental. Furthermore, to predict its possible relevance in other sectors, a detailed study of $Zr_2SeC$ is of scientific importance. Thus, a detailed study of $Zr_2SeC$ needs to be performed to take the full advantages for possible use in many sectors.

Therefore, the structural, electronic, mechanical, thermal, and optical properties of $Zr_2SeC$ have been presented in this paper, and the properties of $Zr_2SeC$ have been compared with those of the $Zr_2SC$ MAX phase. It is found that the $Zr_2SeC$ is soft with low Vickers hardness like other phase materials. Moreover, it is suitable to be used as a TBC material. Furthermore, it can also be used in optical and optoelectronic devices, and as cover material for spacecraft to reduce solar heating. The rest of the article is organized as follows: the detailed computational methodology is given in section 2, results and discussion are presented in section 3, and important conclusions are drawn in section 4.

## 2. Computational methodology

A density functional theory based on the plane-wave pseudopotential method was implemented in the Cambridge Serial Total Energy Package (CASTEP) code [45,46] to calculate the physical properties of $Zr_2SeC$. The exchange and correlation functions were treated by the generalized gradient approximation (GGA) of the Perdew–Burke–Ernzerhof (PBE) [47]. The pseudo-atomic calculations were performed for C - $2s^2\,2p^2$, Se- $4s^2 4p^4$, and Zr - $4d^2\,5s^2$ electronic orbitals. A k-point [48] mesh of size $10 \times 10 \times 3$ was selected to integrate the Brillouin zone, and the cutoff energy was set as 500 eV. The Broyden Fletcher Goldfarb Shanno (BFGS) technique [49] was used for structure relaxation, and density mixing was used for electronic structure calculation.

The structure was relaxed with the following parameters: the self-consistent convergence of the total energy is $5 \times 10^{-6}$ eV/atom, the maximum force on the atom is 0.01 eV/Å, the maximum ionic displacement is set to $5 \times 10^{-4}$ Å, and maximum stress of 0.02 GPa. The optical properties of the Zr$_2$SeCcompoundwere calculated using the complex dielectric function $\varepsilon(\omega) = \varepsilon_1(\omega) + i\varepsilon_2(\omega)$. The time-dependent perturbation theory (first-order) is used to obtain the imaginary part of the dielectric function. The equation used to calculate the imaginary part $\varepsilon_2(\omega)$ is given by:

$$\varepsilon_2(\omega) = \frac{2e^2\pi}{\Omega\varepsilon_0} \sum_{k,v,c} \left|\psi_k^c \left|\mathbf{u.r}\right| \psi_k^v\right|^2 \delta\left(E_k^c - E_k^v - E\right),$$

Where $\Omega$ and $\varepsilon_0$ correspond to the volume of the unit cell and dielectric constant of the free space, **u** and **r** stand for the incident electric field vector (polarized) and position vector. $\omega$, $e$ and $\psi_k^c$ and $\psi_k^v$ stand for the light frequency, electronic charge, and conduction and valence band wave functions at $k$, respectively. Here, the sum $k$ was used to present the Brillouin zone' sampling in the $k$ space. The sums $v$ and $c$ were used to represent the contribution from the unoccupied conduction band (CB) and occupied valence band (VB).The Kramers–Kronig relations were used to estimate the real part $\varepsilon_1(\omega)$ from the imaginary part $\varepsilon_2(\omega)$. The real and imaginary parts of the dielectric function were further used to obtain the other optical constants: refractive index, extinction coefficient, absorption spectrum, reflectivity, and energy-loss spectrum based on the relations found elsewhere [50].

1. **Results and discussion**

**3.1 Structural properties**

The unit of Zr$_2$SeC is shown in Fig. 1, which is crystallized in the hexagonal system with space group P63/mmc (194) [2]. The unit cell contains two formula units, and there are eight atoms in the unit cell. The Zr$_6$C octahedron interleaved between two atomic layers of the Se atom is shown in Fig 1. For 211 MAX phases, the atomic positions of Zr, Se, and C atoms in the unit cell are (1/3, 2/3, 0.0965), (1/3, 2/3, 3/4), and (0, 0, 0). The unit of Zr$_2$SeC is optimized geometrically to calculate the physical properties further. The lattice constants of the optimized cell are presented together with previous experimental and theoretical results [44] that assure the parameters used for calculations. For example, our calculated values of *a*, *c*, and *V* are only 0.1%, 0.18%, and 1.08%, respectively, higher than those of experimental values.

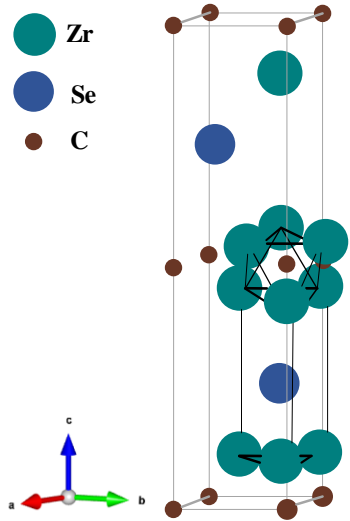

**Table 1 -** Calculated lattice parameters (*a* and *c*), *c/a* ratio, volume (V), and atomic positions of Zr$_2$SeC MAX phase.

| a (Å) | c (Å) | c/a | V (Å$^3$) | Ref. | Positions | Zr | Se | C |
|---|---|---|---|---|---|---|---|---|
| 3.4655 | 12.5406 | 3.618 | 130.429 | *This study* | x | 1/3 | 1/3 | 0 |
| 3.462 | 12.518 | 3.615 | 129.029 | *Expt.* [44] | y | 2/3 | 2/3 | 0 |
| 3.487 | 12.631 | 3.622 | 132.080 | *Theo.* [44] | z | 0.0965 | 3/4 | 0 |
|  |  |  |  |  |  | 0.0963 [44] |  |  |

## *3.2 Electronic properties and bonding nature*

The electronic band structure of Zr$_2$SeC of optimized structure within the GGA-PBE is estimated and analyzed as shown in Fig. 2 (a) along the high symmetry lines of the Brillouin zone (Γ–A–H–K–Γ–M–L–H). None existence of a bandgap close to the Fermi level owing to the overlapping of the conduction and valence band reveals the metallic nature of the titled compound like other MAX phases [51–53]. The band structure also reveals the anisotropic nature of electronic conductivity. It is seen that the energy dispersion along Γ–A, K–H, and L–M directions, which are along the *c*-direction, are small. The conductivity in the basal plane is exhibited by H–K, Γ–M, and L–H that are larger than those of *c*-direction. Thus, the conductivity is higher in the basal plane than that of *c*-direction for Zr$_2$SeC, which is typical for MAX phases [21,30,52].

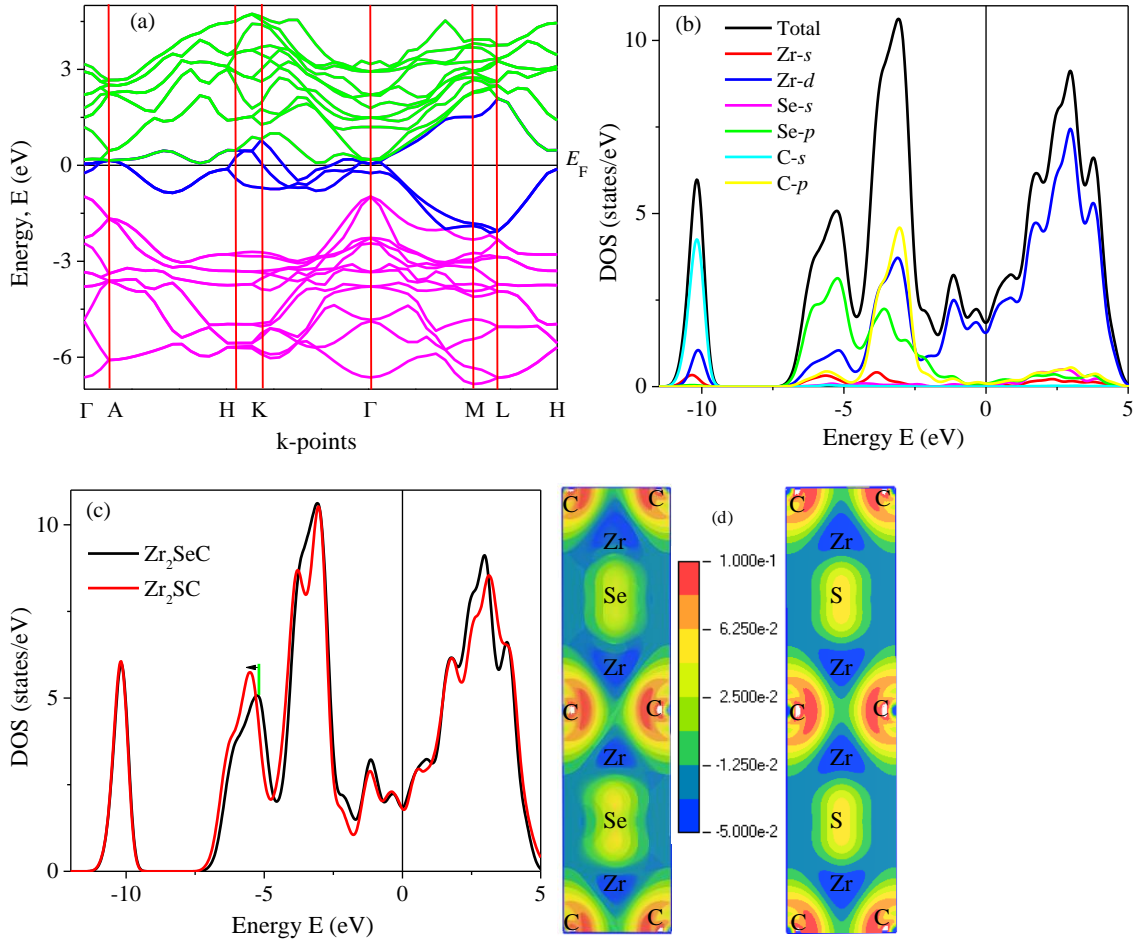

After band structure, the next step for insight into the electronic properties of $Zr_2SeC$ is to investigate its density of states (DOS). Fig. 2 (b) shows the DOS of $Zr_2SeC$ with a finite value of DOS (1.84 states per unit cell per eV) at the Fermi level that is slightly higher than the value (1.48 states per unit cell per eV) obtained by Chen et al.[44], might due to the difference in the functionals used. The DOS of $Zr_2SeC$ is slightly higher than that of $Zr_2SC$ (1.79 states per unit cell per eV); a similar result is also reported by Chen et al.[44]. The value of DOS at the Fermi level ($E_F$) can provide information regarding electrical conductivity as it is directly related to the DOS and electron mobility [2]. The partial DOS is used to explore the contribution from different states to the total DOS. For instance, the DOS at the $E_F$ is contributed from Zr-$d$ states and involved in the conduction of $Zr_2SeC$. The C and Se do not contribute to the DOS at $E_F$ and are not involved in the conduction properties. The contribution of different atoms to the DOS at different energy states is also observed from the partial DOS (PDOS). The DOS profile is almost similar to that of the reported DOS [44]. The hybridization among the different Zr, Se, and C electronic states is observed from PDOS. The lowest energy band (- 9 to - 12 eV) comes from the hybridization between Zr-*d* and C-*s* orbital's electrons with a dominating role of C-*s* states. The peak in the energy range -4.5 to -7.5 eV is accredited by the strong hybridization of Se-*p* and Zr-*d* states with a minor contribution from Zr-*s* and C-*p* states. The strongest peak in the range – 1.7 to 4.5 eV results from the hybridization between Zr-*d*, Se-*p*, and C-*p* states with dominant contribution from C-*p* states.

The highest valence band in the range 0 to 1.7 eV is dominated by Zr-*d* states with a small contribution from Se-*p* and C-*p* states. The PDOS of $Zr_2SC$ is not presented here due to similar nature. The TDOS of $Zr_2AC$ (A = S, Se) is shown in Fig. 2 (c) to disclose the effect of the A element. As seen in Fig 2 (b), the S/Se contributed to the TDOS mostly in the energy range of -4.5 to -7.5 eV. The peak found in this region is shifted towards lower energy (indicated by the green line and black arrow) for $Zr_2SC$ compared to $Zr_2SeC$, indicating a higher bonding M-A strength for $Zr_2SC$ compared to that in $Zr_2SeC$. The peaks position due to Zr and C remain the same for both compounds. Thus, higher values of the mechanical properties characterizing parameters are expected owing to the presence of stronger M-A bonding in $Zr_2SC$ than in $Zr_2SeC$. To strengthen this statement, we have calculated the charge density mapping (CDM) for both phases, as shown in Fig 2(d). Figure 2 (d) shows the CDM for both $Zr_2AC$ (A = S, Se), where the red and blue colors indicate the highest and the lowest value, respectively. The directional covalent bonding occurs between Zr and C atoms where the charges are geometrically localized, and this boding is comparatively stronger, required more energy to break this bonding. As seen in the Figure, the charge density at Zr and C position is unchanged for both$Zr_2SC$ and $Zr_2SeC$. The charges are accumulated to a greater extent at S positions of $Zr_2SC$ than that of Se positions of $Zr_2SeC$. The greater charges at the S position the greater bonding strength between Zr - S than Zr – Se which is in good agreement with DOS results. Now it's time to prove the statement that the Zr- S bonding stronger than Zr – Se which will be proven in the following section.

*3.3 Mechanical properties*

Some of the physical properties of solids, for example, the mechanical stability, bonding strength, deformation, failure mode, stiffness, anisotropic nature in bonding strength, etc., can be brought out by studying mechanical properties characterizing parameters that are subjected in this section. At first, the stiffness constants ($C_{ij}$) have been calculated using the well-known strain-stress method[54–59].The stiffness constants, which are five in number as independent due to the hexagonal nature of Zr$_2$SeC and presented in Table 2 together with those of other S-containing MAX phases Zr$_2$SX (x = C, B). One of the prime importance of stiffness constants is the use of checking the mechanical stability of solids. Max Born [60]proposed some conditions on the stiffness constants of solids for being mechanically stable, having a limitation that they are not enough to predict the stability for all crystal systems correctly. The limitation is overcome by Mouhat et al. [61]and the stability conditions for a hexagonal systems becomes: $C_{11} > 0$, $C_{11} > C_{12}$, $C_{44} > 0$, $(C_{11} + C_{12})C_{33} - 2(C_{13})^2 > 0$. The $C_{ij}$ presented in Table 2 satisfy the above relations, and hence Zr$_2$SeC is expected to be mechanically stable like M$_2$SC (M =Zr, Hf, Nb). It should be noted here that Zr$_2$SeC is already experimentally realized; thus, the question regarding stability is not expected at all. But, checking mechanical stability is necessary for practical application under load. The importance of $C_{ij}$ is not only limited to mechanical stability checking but also provides significant information. For example, the stiffness of solids along [100] and [001] directions are defined by $C_{11}$ and $C_{33}$, respectively, thus, comparatively low pressure is required to deform Zr$_2$SeCalong crystallographic *a*-axis than *c*-axis because of $C_{11} < C_{33}$. The resistance to shear deformation is measured by the value of $C_{44}$. It is evident from Table 2 that $C_{44} < C_{33}$ and $C_{11}$, indicating a low pressure required to shear deformation than axial deformation. Moreover,$C_{44}$is also a hardness predictor, in fact, directly related to the hardness compared to other elastic moduli [62]. $C_{44}$of Zr$_2$SeC is smaller than M$_2$SC (M = Zr, Hf) but larger than Nb$_2$SC, thus, the hardness of Zr$_2$SeCis expected to be lower than M$_2$SC (M = Zr, Hf) but higher than Nb$_2$SC. Furthermore, the unequal values of $C_{11}$ and $C_{33}$reveal the anisotropy in the bonding strength in the *a* and *c*-axis owing to the difference in the atomic arrangement in *a* and *c*-axes. The difference in the atomic arrangement is mainly responsible for the mechanical anisotropy that will be presented in the letter.

**Table 2:**The elastic constants, $C_{ij}$ (GPa), bulk modulus, *B* (GPa), shear modulus, *G* (GPa), Young's modulus, *Y* (GPa), macro, $H_{macro}$ (GPa), micro hardness, $H_{micro}$ (GPa), Pugh ratio, *G/B*, Poisson ratio, ν and Cauchy Pressure, *CP* (GPa) of Zr$_2$SeC, together with those of M$_2$SC (M = Zr, Hf, Nb).

| Phase | $C_{11}$ | $C_{12}$ | $C_{13}$ | $C_{33}$ | $C_{44}$ | B | G | Y | $H_{macro}$ | $H_{micro}$ | G/B | ν | CP | Reference |
|---|---|---|---|---|---|---|---|---|---|---|---|---|---|---|
| Zr$_2$SeC | 260 | 97 | 96 | 293 | 128 | 154 | 100 | 247 | 14.85 | 17.79 | 0.65 | 0.23 | -30 | Thisstudy |
| Zr$_2$SC | 295 | 89 | 102 | 315 | 138 | 166 | 115 | 280 | 17.89 | 21.57 | 0.69 | 0.22 | -49 | [35] |
| Hf$_2$SC | 311 | 97 | 121 | 327 | 149 | 181 | 120 | 295 | 17.35 | 21.72 | 0.66 | 0.23 | -52 | [35] |

| Nb$_2$SC | 316 | 108 | 151 | 325 | 124 | 197 | 105 | 267 | 11.58 | 15.84 | 0.53 | 0.27 | -16 | [35] |

Bulk modulus (*B*) defines the property related to the incompressibility of the materials. The material's resistance to change its volume keeping shape unchanged under hydrostatic pressure is measured by its *B*. The resistance of solids to plastic deformation at constant volume is provided by its shear modulus (*G*). The stiffness of solids is measured by its Young's modulus (*Y*); the higher value of *Y* indicates stiffer solids and vice versa. These moduli are also important to realize the hardness, stiffness, brittleness/ductileness of solids. From these points of view, we have calculated the *B* and *G* using Hill's approximation [63,64] based on the Voigt [65] and Reuss [66] models as follows: $B = (B_V + B_R)/2$, where $B_V = [2(C_{11} + C_{12}) + C_{33} + 4C_{13}]/9$ and $B_R = C^2/M$; $C^2 = C_{11} + C_{12})C_{33} - 2C_{13}^2$; $M = C_{11} + C_{12} + 2C_{33} - 4C_{13}$. $B_V$ represents the upper limit of *B*(Voigt bulk modulus) and $B_R$ represents the lower limit of *B* (Reuss bulk modulus). Like *B*, average values of Voigt ($G_V$) and Reuss ($G_R$) were used to calculate *G* using the following equations: $G = (G_V + G_R)/2$, where $G_V = [M + 12C_{44} + 12C_{66}]/30$ and $G_R = \left(\frac{5}{2}\right)[C^2 C_{44}C_{66}]/[3B_V C_{44}C_{66} + C^2(C_{44} + C_{66})]$; $C_{66} = (C_{11} - C_{12})/2$. The Young's modulus (*Y*) and Poisson's ratio (*υ*) were also calculated using their relationships with *B* and *G*: $Y = 9BG/(3B + G)$ and $υ = (3B - Y)/(6B)$[67,68]. The lower values of *B* and *G* of Zr$_2$SeC than those of Zr$_2$SC revealed that low pressure is required to volume and plastic deformation for Zr$_2$SeC than Zr$_2$SC and lower value of *Y* for Zr$_2$SeC than Zr$_2$SC, indicating that Zr$_2$SeC is less stiff than Zr$_2$SC. The elastic moduli were further used to study the hardness of Zr$_2$SeC using the equations: $H_{macro} = 2(k^2 G)^{0.585} - 3$ (where Pugh's ratio, *k=G/B*) [69] and $H_{micro} = \frac{(1-2v)E}{6(1+v)}$ [70]. The study of the hardness parameter is beneficial for those materials which are used as structural components like MAX phases. The replacement of Se by S is enough to increase the M-A bonding strength that must improve the hardness of Zr$_2$SC than Zr$_2$SeC. The obtained values of $H_{macro}$ and $H_{micro}$ = 14.85 (17.89) and 17.79 (21.57) GPa for Zr$_2$SeC (Zr$_2$SC). Thus, the enhancement of hardness for Zr$_2$SC is observed. Though the elastic moduli (*B, G, Y*) ofZr$_2$SeC are smaller than that of Nb$_2$SC but the hardness parameters of Zr$_2$SeC are larger than those of Nb$_2$SC owing to the larger value of $C_{44}$[62] ($C_{44}$of Zr$_2$SeC and Nb$_2$SC are 128 and 124 GPa, respectively).The elastic moduli and hardness parameters of Hf$_2$SC are larger than those of Zr$_2$SeC. The lowering of the bonds strength (Zr-C and Zr- Se bond) for Zr$_2$SeC compared to Zr$_2$SC (Zr-C and Zr- S bond) is responsible for the lowering of elastic moduli and hardness for the same. The bond length of Zr-C (2.33869 Å) in $Zr_2SeC$ is higher than in $Zr_2SC$ (2.32604 Å), and Zr-Se (2.77595 Å) in $Zr_2SeC$ is higher than Zr-S in $Zr_2SC$ (2.68671 Å) and hence the bonds (Zr-C, Zr-S) of $Zr_2SC$ are stronger than those of $Zr_2SeC$ (Zr-C, Zr-Se). Thus, lower hardness parameters for $Zr_2SeC$ compared to $Zr_2SC$ is expected as reflected from the hardness parameters presented in Table 2. To be more confirmed, we have calculated the Vickers hardness ($H_v$) using the Gou et al. [71] formula that is based on the Mulliken bond population and geometrical averages of the bonds present within the crystal. The relevant formula can be found elsewhere [72–74]. The calculated values of $H_v$ for $Zr_2SeC$ are $Zr_2SC$ 2.52 and 4.33 GPa, respectively. Again, the Vickers hardness of $Zr_2SeC$ is lower than that of $Zr_2SC$ as expected. The obtained elastic moduli and hardness are in good agreement with the DOS and CDM results.

## 3.4 Mechanical anisotropy

For hexagonal crystal, the elastic anisotropy is related to microcracks and anisotropic plastic deformation, which plays a vital role in understanding the mechanical stability of the material under service. The mechanical anisotropy in MAX phases is normally due to unequal elastic constants, i.e. $C_{11} \neq C_{33}$, which indicates that the mechanical properties in all crystallographic planes are not identical [75]. Moreover, the knowledge of anisotropy offers the information necessary to enhance further the stability of solids for many applications [76]. Therefore, it is necessary to study the anisotropic behavior of mechanical properties in $Zr_2SeC$. For this purpose, the 2D and 3D visualization of Young's modulus, compressibility, shear modulus, and Poisson's ratio of $Zr_2SeC$ MAX phase is presented in Fig. 3 (a-d) in comparison with $Zr_2SC$ [Fig. (a-d)] by using the open-source software packages ELATE and AnisoVis [77,78]. The obtained results for $Zr_2SeC$ and $Zr_2SC$ are tabulated in Table 4. The unit value of the anisotropic index indicates that the material has identical mechanical properties in all directions. The extent of anisotropy is estimated by the variable value of elastic properties in all directions. The 2D projections and 3D plots are perfectly circular and spherical for isotropic materials, and variation from circular/spherical shape demonstrates the degree of anisotropy of the corresponding elastic property. The Young's modulus (*Y*) for $Zr_2SeC$ and $Zr_2SC$ exhibit the isotropic nature in the *xy* plane because 2D planer projections are perfectly circular for both MAX phases [Fig.3 (a) and 4(a)]. The anisotropic behavior is found in the *xz* and *yz* planes because the 2D projection for *Y* has minimum values in the vertical and horizontal axis and the maximum value at around 45° of $z$-axis in the $xz$ and $yz$ planes. The minimum (maximum) values of Y are 208.93 GPa (277.50 GPa) and 250.08 GPa (306.39 GPa) for $Zr_2SeC$ and $Zr_2SC$, respectively. This implies that $Zr_2SC$ shows more anisotropy as compared to $Zr_2SeC$. The color difference in the 3D plots of $Y$ represents the anisotropic behavior and minimum (maximum) values.

Similar to $Y$, the linear compressibility ($K$) of $Zr_2SeC$ and $Zr_2SC$ exhibits isotropic nature in the $xy$ plane and anisotropic nature in the $xz$ and $yz$ planes [(Fig.3 (b) and 4 (b)]. The $Zr_2SeC$ and $Zr_2SC$ show minimum value of 1.91 $TPa^{-1}$ (maximum = 2.28 $TPa^{-1}$) and 1.78 $TPa^{-1}$ (maximum = 2.12 $TPa^{-1}$) values of $K$ in the vertical (horizontal) axis in $xz$ and $yz$ planes, respectively. The degree of anisotropy for $K$ in $Zr_2SeC$ is greater than that of $Zr_2SC$. Fig.3(c) and 4 (c) represent the anisotropy in shear modulus ($G$) of $Zr_2SeC$ and $Zr_2SC$ MAX phases. The 2D projections of $G$ consist of two surfaces in which blue and green lines represent the maximum and minimum values at each particular angle. It is found that the maximum values of $G$ for $Zr_2SeC$ and $Zr_2SC$ MAX phases are at the horizontal and vertical axis in the $xz$ and $yz$ planes and minimum values are at around the 45° of both axes. It is worth mentioning that the minimum value that coincides with the green line shifts towards the horizontal axis in both $xz$ and $yz$ planes. This implies that both MAX phases are anisotropic in $xz$ and $yz$ planes and are likely to be isotropic in the $xy$ plane. The anisotropy of Poisson's ratio ($\upsilon$) is shown in Fig 3 (d) and 4 (d) for $Zr_2SeC$ and $Zr_2SC$. Like $G$, there are two lines: blue and green, which indicate the maximum and minimum values for each angle. The isotropic (in the $xy$ plane) and anisotropic (in the $xz$ and $yz$ planes) nature of $\upsilon$ can be seen from the 2D projections. Moreover, the maximum and minimum values of mechanical parameters for $Zr_2SeC$ and $Zr_2SC$ are listed in Table 4. It is found that the $Zr_2SeC$ is qualitatively more anisotropic as compared to $Zr_2SC$.

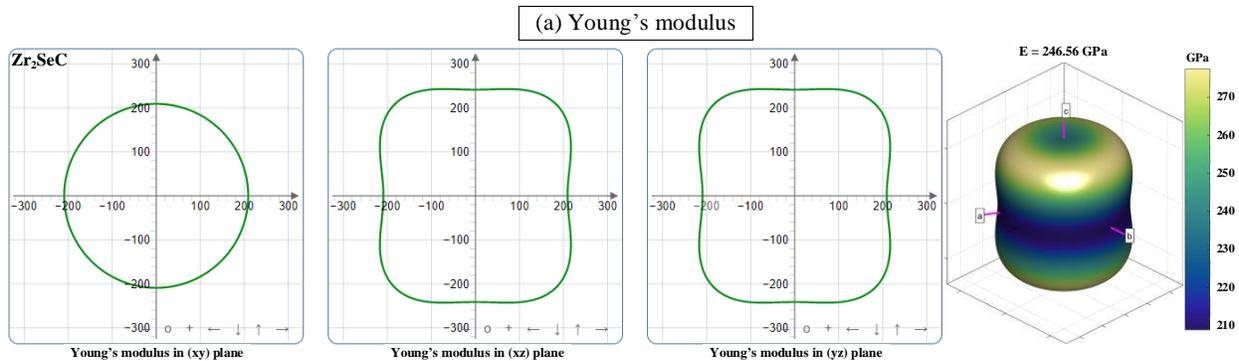

(a) Young's modulus

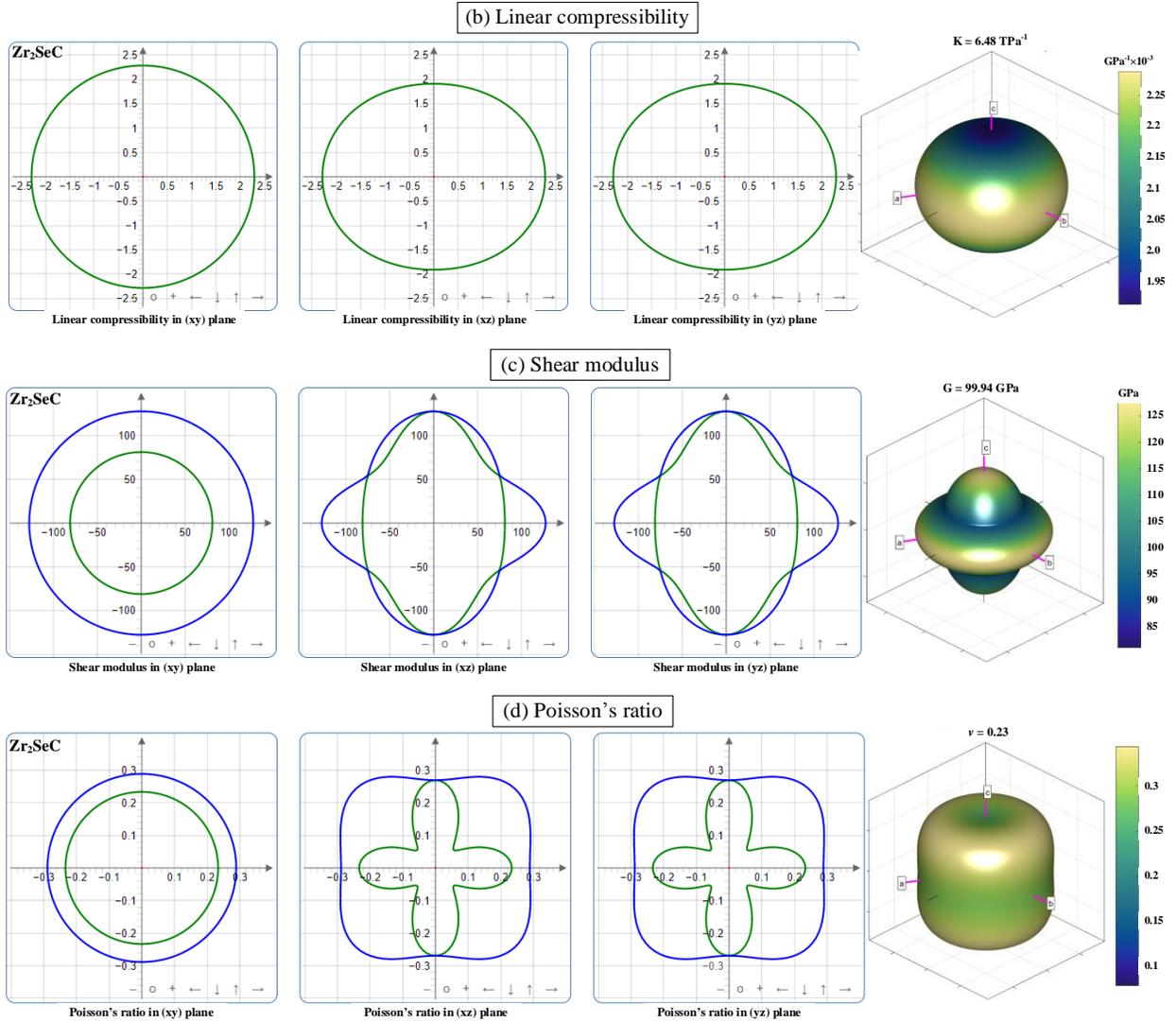

Fig. 3 The 2D projections and 3D plots of (a) Young's modulus, Y, (b) Linear compressibility, K, (c) shear modulus, G, and (d) Poisson's ratio, ν of Zr$_2$SeC MAX phase.

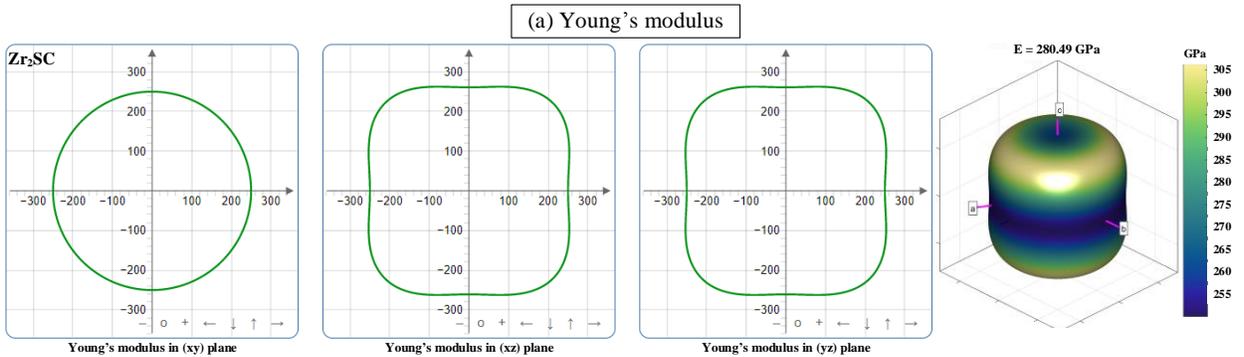

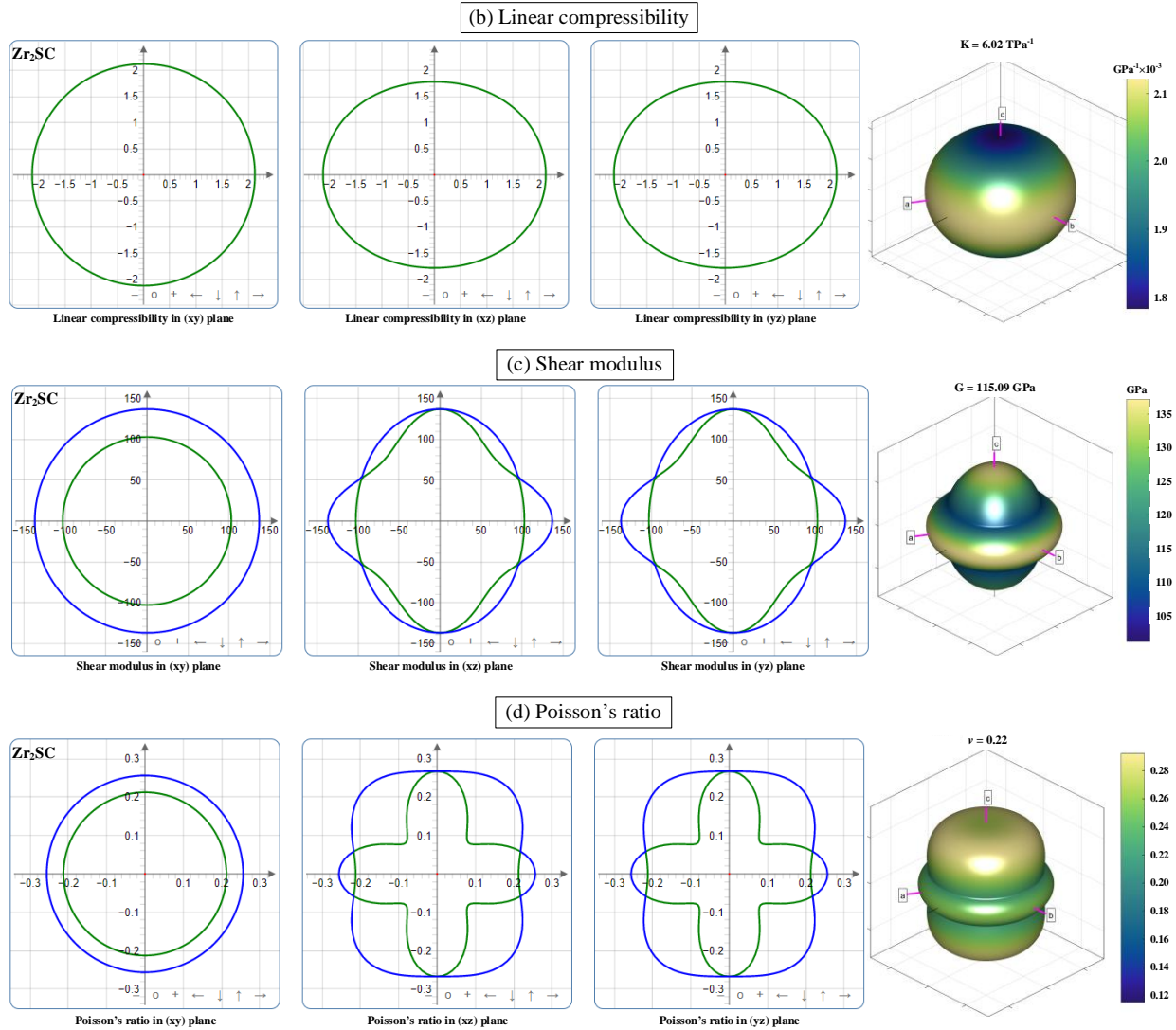

Fig. 4 The 2D projections and 3D plots of (a) Young's modulus, Y, (b) Linear compressibility, K, (c) shear modulus, G, and (d) Poisson's ratio, $v$ of Zr$_2$SC MAX phase.

Table – 4 The minimum and maximum values of Young's modulus, Y (GPa), linear compressibility, K (TPa$^{-1}$), shear modulus, G (GPa), and Poisson's ratio, $v$ and their anisotropic indices, A of Zr$_2$AC (A = Se, S).

| Phases | Y$_{min}$ | Y$_{max}$ | A$_Y$ | K$_{min}$ | K$_{max}$ | A$_K$ | G$_{min}$ | G$_{max}$ | A$_G$ | $v_{min}$ | $v_{max}$ | A$v$ | Ref. |
|---|---|---|---|---|---|---|---|---|---|---|---|---|---|
| Zr$_2$SeC | 208.93 | 277.50 | 1.32 | 1.913 | 2.288 | 1.19 | 81.071 | 127.59 | 1.57 | 0.078 | 0.343 | 4.394 | This work |
| Zr$_2$SC | 250.08 | 306.39 | 1.22 | 1.78 | 2.12 | 1.19 | 101.19 | 137.27 | 1.35 | 0.110 | 0.290 | 2.53 | [35] |

We have also investigated the different anisotropic factors for the {100}, {010} and {001} planes that are computed using the equations: $A_1 = \frac{\frac{1}{6}(C_{11}+C_{12}+2C_{33}-4C_{13})}{C_{44}}$, $A_2 = \frac{2C_{44}}{C_{11}-C_{12}}$, $A_3 = A_1 \cdot A_2 = \frac{\frac{1}{3}(C_{11}+C_{12}+2C_{33}-4C_{13})}{C_{11}-C_{12}}$[79], respectively and presented in Table 5. Since the values of $A_i$'s are not equal to 1 (one), thus $Zr_2SeC$ and $Zr_2SC$ are anisotropic because $A_i = 1$ implies the isotropic nature. The bulk modulus for $a$ and $c$-direction are computed using the equations[80]: $B_a = a\frac{dP}{da} = \frac{\Lambda}{2+\alpha}, B_c = c\frac{dP}{dc} = \frac{B_a}{\alpha}$, where $\Lambda = 2(C_{11} + C_{12}) + 4C_{13}\alpha + C_{33}\alpha^2$ and $\alpha = \frac{(C_{11}+C_{12})-2C_{13}}{C_{33}+C_{13}}$. The unequal values of $B_a$ and $B_c$ [Table. 5] indicates the anisotropic nature of $Zr_2AC$ (A = Se, S). The linear compressibility ($k$) along the $a$ and $c$-axis are calculated by the equation [81]: $\frac{k_c}{k_a} = C_{11} + C_{12} - 2C_{13}/(C_{33} - C_{13})$. The obtained values of $k_c/k_a$ are not equal to 1 ($k_c/k_a = 1$ for isotropic materials) and hence $Zr_2AC$ (A = Se, S) are anisotropic. Another important anisotropic factor, the universal anisotropic index $A^U$ is estimated by the following equation [82]: $A^U = 5\frac{G_V}{G_R} + \frac{B_V}{B_R} - 6 \geq 0$ where $B$ and $G$ are obtained by Voigt and Reuss models. Since, the values of $A^U$ are greater than zero, indicating the anisotropic nature of $Zr_2AC$ (A = Se, S).

**Table 5.** The anisotropic factors, $A_1$, $A_2$, $A_3$, $B_a$, $B_c$, $k_c/k_a$, and universal anisotropic index $A^U$ of $Zr_2AC$ (A = Se, S).

| Phase | $A_1$ | $A_2$ | $A_3$ | $k_c/k_a$ | $B_a$ | $B_c$ | $A^U$ |
|---|---|---|---|---|---|---|---|
| $Zr_2SeC$ | 0.73 | 1.57 | 1.14 | 0.84 | 383 | 904 | 0.218 |
| $Zr_2SC$ | 0.73 | 1.34 | 0.98 | 0.85 | 412 | 955 | 0.113 |

*3.5 Thermal properties*

The thermodynamic properties of the material at high temperatures and pressures are of scientific and technical significance, which help to predict the material's applications at elevated temperatures and pressure. Here we have investigated the thermodynamic properties of the newly synthesized $Zr_2SeC$ MAX phase in comparison with $Zr_2SC$ over the wide temperature (from 0 to 1600 K) and pressure (from 0 to 50 GPa) by using the quasi-harmonic Debye approximation [83,84]. To disclose the important thermodynamic properties, the thermal expansion coefficient(*TEC*), Debye temperature ($\theta_D$), entropy (S), heat capacity at constant volume ($C_v$), Grüneisen parameter ($\gamma$) along with volume (V) and Gibbs free energy (G) have been investigated in this temperature and pressure range. The quasi-harmonic Debye model remains valid in this given range of temperature and pressure and has been successfully used to calculate the thermodynamic properties of MAX phases [85,86]. The temperature dependence of volume for $Zr_2SeC$ and $Zr_2SC$ MAX phases is depicted in Fig 5. It is found that the volume of both MAX phases expands linearly with an increase in temperature while contracts during the increase in pressure. It is also found that the effect of pressure in volume contraction is more significant than the effect of temperature in volume expansion. Moreover, the volume is constant up to the T ≤ 300 K and begins to increase as the temperature is increased from 300 K. It is also observed that the volume of $Zr_2SeC$ is greater than that of $Zr_2SC$ at any constant temperature and pressures the volume at 300 K for the both MAX phases are with good agreement with the optimized values obtained from the first-principles calculations [Table 1].

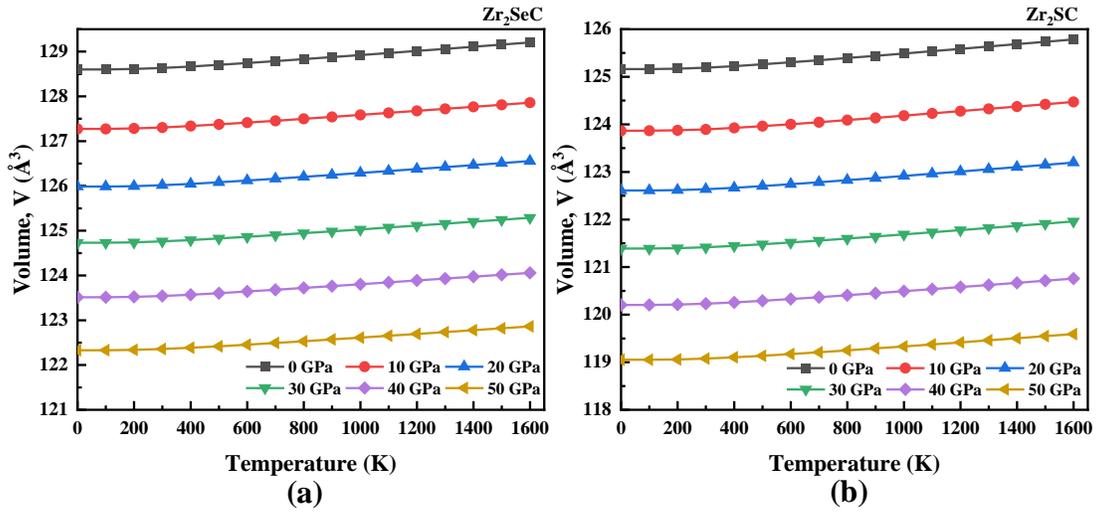

Fig.5 The temperature effect on the volume of (a) $Zr_2SeC$ and (b) $Zr_2SC$ at different pressure.

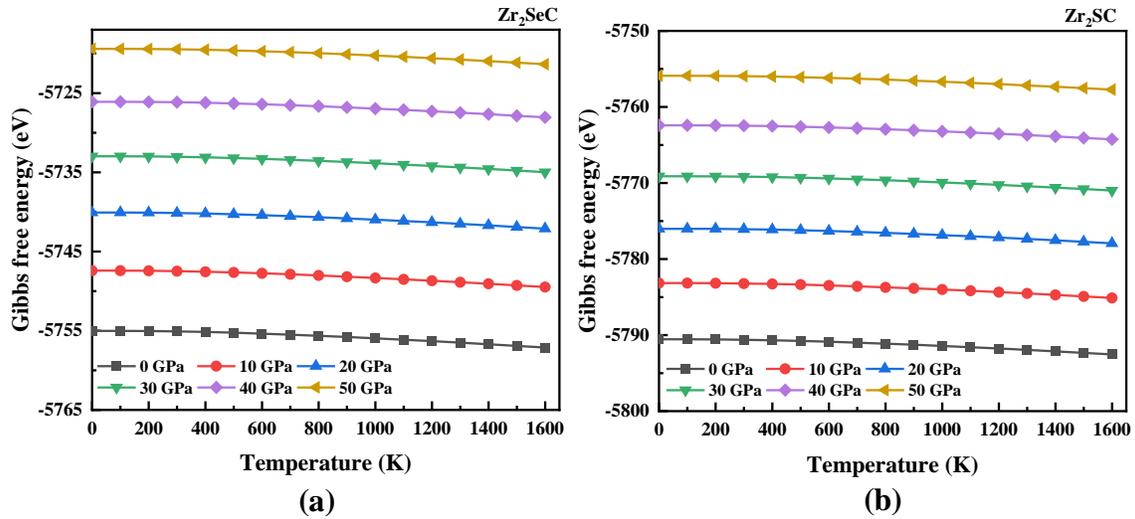

Fig. 6 The variation of Gibbs free energy with temperature for (a) $Zr_2SeC$ and (b) $Zr_2SC$ at different pressure.

Fig. 6 shows the effect of temperature on Gibbs free energy of $Zr_2SeC$ and $Zr_2SC$. It is noticed that the Gibbs free energy decreases linearly to more negative values with the increase in temperature while increases with an increase in pressure. Fig. 7 represents the Variation of Debye temperature ($\theta_D$) of $Zr_2SeC$ and $Zr_2SC$ with the temperature and pressure. It is found that the $\theta_D$ for $Zr_2SeC$ is smaller than that of $Zr_2SC$, and the calculated value for $Zr_2SeC$ and $Zr_2SC$ at zero pressure and T = 300 K are 699.9 K and 764.2 K. The Debye temperature of material plays a vital role in many physical properties such as thermal expansion coefficient and specific heat, and also can evaluate the bonding strength in the solids. The atomic vibrations increase with an increase in temperature, which leads to a decrease in bonding strength; consequently, the Debye temperature decreases. On the other hand, increasing the pressure strengthens the bonding forces between the atoms, and as a result, Debye temperature increases. It is found that the Debye temperature for both MAX phases slowly decreases with the increase in temperature, while the pressure effect is opposite to the temperature because it increases rapidly with the increase in pressure. It is worth mentioning that the Debye temperature is pressure sensitive as compared to temperature because the increasing rate with pressure is larger than that of decreasing rate with temperature. The temperature dependence of the thermal expansion coefficient (*TEC*) is displayed in Fig 8. It is seen that the *TEC* increases rapidly with the increase in temperature up to 300 K. The increase in *TEC* of $Zr_2SeC$ and $Zr_2SC$ becomes less sensitive of temperature at the T $\geq$ 300 K. However, at the constant temperature, *TEC* decreases with the increase in pressure. The calculated value of *TEC* for $Zr_2SeC$ = $2.81\times10^{-6}$ $K^{-1}$ < $Zr_2SC$ = $2.96\times10^{-6}$ $K^{-1}$ at zero pressure and T = 300 K.

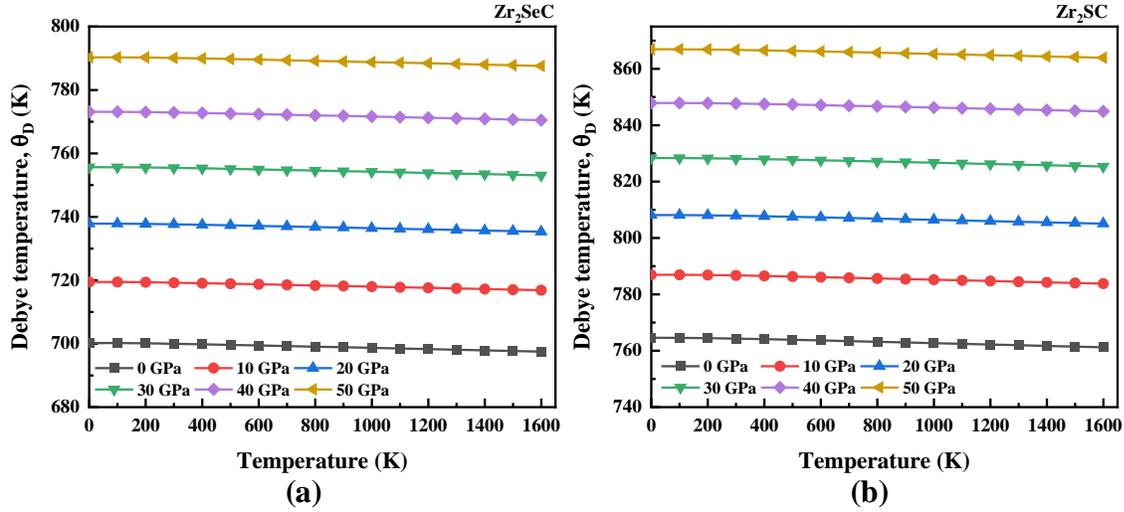

Fig. 7 The temperature effect on the Debye temperature of (a) $Zr_2SeC$ and (b) $Zr_2SC$ at different pressure.

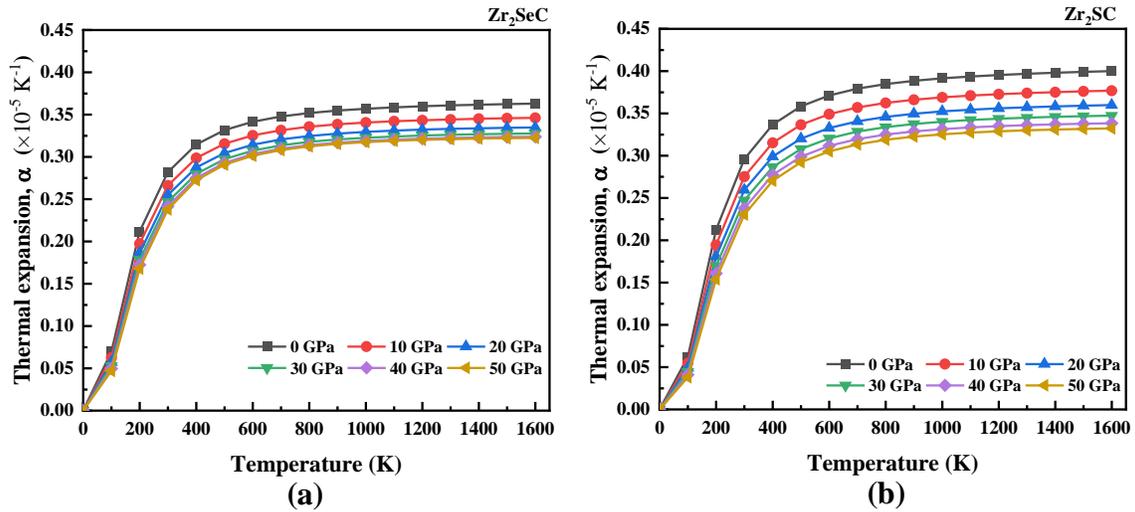

Fig.8 The temperature effect on the thermal expansion coefficient of (a) $Zr_2SeC$ and (b) $Zr_2SC$ at different pressure.

The specific heat is significant to predict the density of states, band structure, and lattice vibrations. The effect of temperature on the specific heat at constant volume ($C_v$) for $Zr_2SeC$ and $Zr_2SC$ is given in Fig 9. The heat capacity for both MAX phases satisfies the Debye $T^3$ power-law at the low temperatures, which is mainly due to an exponential increase in the number of exciting phonon modes. At high temperatures, the specific heat curves become converged to get the classical Dulong-petit (DP) limit ($C_v = 3nN_Ak_B$). The DP limit for $Zr_2SeC$ and $Zr_2SC$ is found to be at ~98.83 $Jmol^{-1}K^{-1}$ and ~98.65$Jmol^{-1}K^{-1}$. The heat capacity decreases linearly with the increase in pressure. The calculated value of $C_v$ for $Zr_2SeC$ (77.13 Jmol$^{-1}$K$^{-1}$) is greater than that of $Zr_2SC$ (73.63Jmol$^{-1}$K$^{-1}$) at zero pressure, and T = 300 K. It is also observed in Fig. 9 that the heat capacity of $Zr_2SeC$ reached the DP limit at a lower temperature compared to $Zr_2SC$. The values $\Theta_D$ and $H_v$ of $Zr_2SeC$ are smaller than those of $Zr_2SC$;thus, it is expected to reach the DP limit at a lower temperature for $Zr_2SeC$ than that of $Zr_2SC$. A similar result is also found for other solids [57,87].

Fig. 10 represents the entropy (S) of $Zr_2SeC$ and $Zr_2SC$ as a function of temperature at different pressure. It is found that the entropy increases with an increase in temperature and decreases with the increase in pressure. At lower temperatures, the effect of pressure on entropy is almost negligible. However, the entropy increases exponentially as the temperature is above 100 K. The entropy is more sensitive to temperature than the pressure in the given range of temperature and pressure.

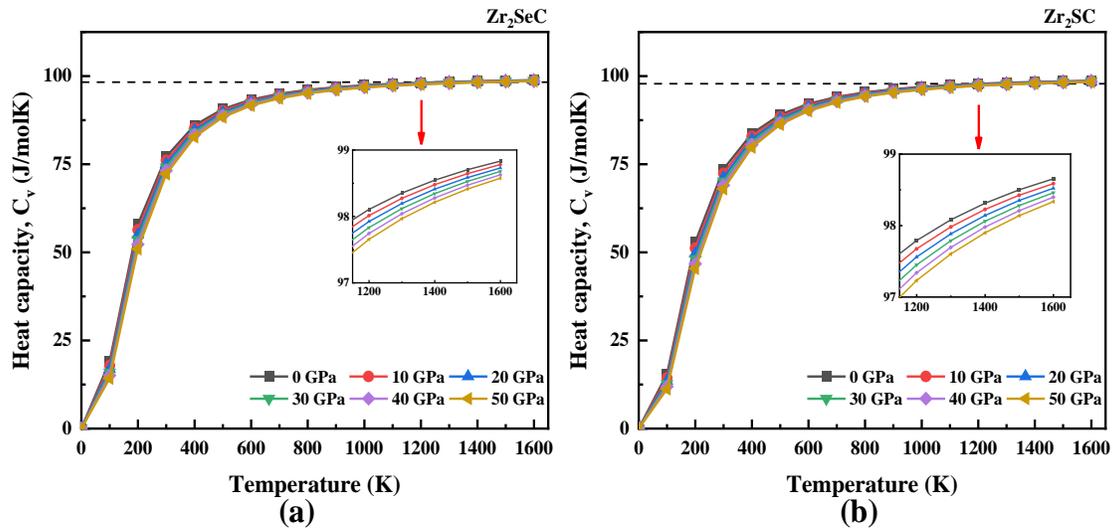

Fig. 9 The temperature on the heat capacity at constant volume, $C_v$ of (a) $Zr_2SeC$ and (b) $Zr_2SC$ at different pressure.

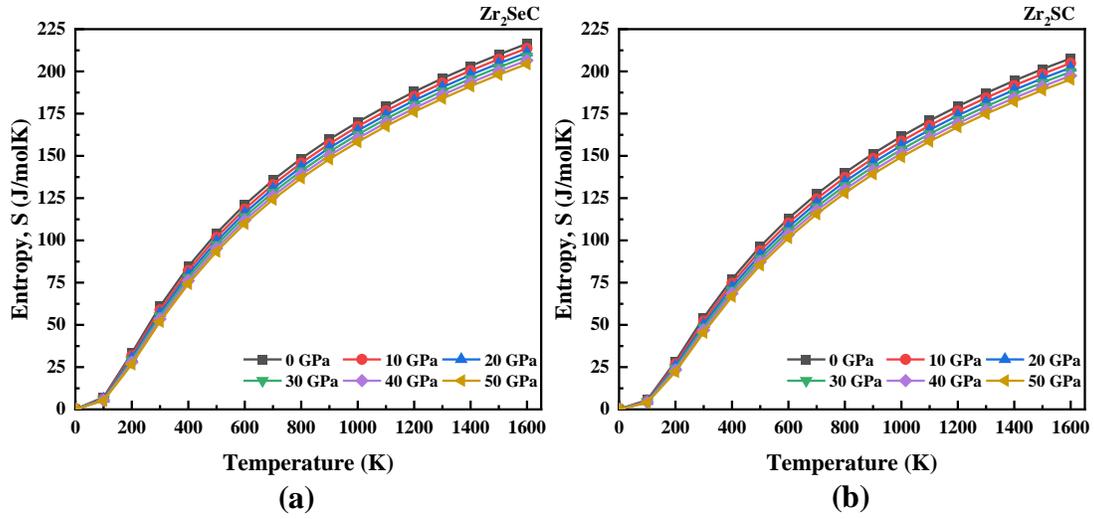

Fig. 10 The temperature effect on the entropy of (a) $Zr_2SeC$ and (b) $Zr_2SC$ at different pressure.

The Grüneisen parameter (γ) for $Zr_2SeC$ and $Zr_2SC$ as a function of temperature and pressure is depicted in Fig 11. It is an important parameter that is used to calculate the thermal states and can also quantify the anharmonic effect. The Grüneisen parameter for $Zr_2SeC$ and $Zr_2SC$ increases with the increase in temperature at lower pressures and decreases with an increase in temperature at higher pressures, as shown in Fig. 11 (a and b). The effect of pressure on γ for $Zr_2SeC$ and $Zr_2SC$ is shown in Fig 11 (c and d). For $Zr_2SeC$, the γ decreases with an increase in pressure and becomes constant at 20 GPa and then increases rapidly at higher pressures. Similarly, the γ for $Zr_2SC$ decreases linearly with the increase in pressure and becomes constant at 30 GPa and then increases linearly with the increase in pressure. The calculated value of γ for $Zr_2SeC$ (0.755) is smaller than that of $Zr_2SC$ (0.806) at zero pressure and T = 300 K.

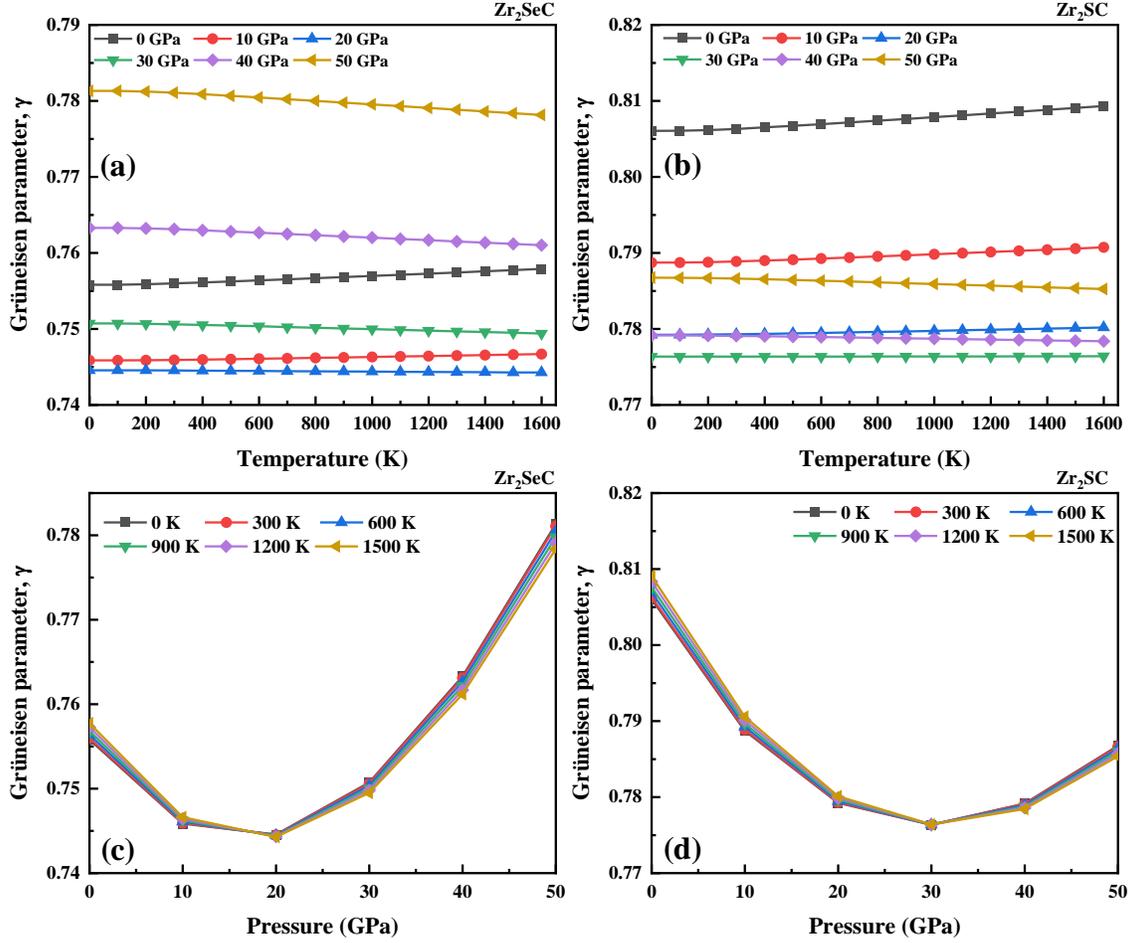

Fig. 11 The effect of temperature and pressure on the Grüneisen parameter of (a, c) $Zr_2SeC$ and (b, d) $Zr_2SC$, respectively.

Moreover, we have also calculated two important parameters regarding the application at high temperatures, such as the minimum thermal conductivity and melting temperature. The thermal conductivity of the materials which are used at high temperatures (e.g., ceramics) reduces to the smallest value ($K_{min}$) at high temperatures. The $K_{min}$ of $Zr_2AC$ (A = Se, S) is computed by the equation [88]: $K_{min} = k_B v_m \left(\frac{M}{n\rho N_A}\right)^{-2/3}$, where $k_B$, $v_m$, $N_A$ and $\rho$ are Boltzmann constant, average phonon velocity, Avogadro's number and density of crystal, respectively. The obtained values of $K_{min}$ are 1.3 and 1.6 (W/mK) for $Zr_2SeC$ and $Zr_2SC$, respectively which are comparable with those of other MAX phase materials [7,30,31]. The thermal conductivity of $Zr_2SeC$ (18.30 W/mK) is also lower than that of $Zr_2SC$ (21.10 W/mK) as measured by Chen et al. [44]. They have also measured the electronic contribution to the electrical conductivity, which is important for predicting high-temperature application. Accordingly, the electronic contribution to the thermal conductivity is 25 % for $Zr_2SeC$ and 19.3 % for $Zr_2SC$; thermal conductivity is dominated by phonon contribution. Thus, $K_{min}$ of $Zr_2SeC$ and $Zr_2SC$ is also dominated by phonon contribution and the low values of $K_{min}$ could be useful to predict its suitability as thermal barrier coating (TBC) material. Elastic constants can be used to calculate the melting temperature ($T_m$) of $Zr_2SeC$ and $Zr_2SC$ by using the expression [89]: $T_m = 3C_{11}+1.5C_{33}+354$. It is found that the melting temperature of $Zr_2SeC$ (1571 K) is less than that of $Zr_2SC$ (1714 K), which is comparable with the melting point of $Zr_2GaC$ (1481 K), $Hf_2GaC$ (1648 K), and $Hf_3SnC_2$ (1773 K)[59,75]. Recently, the research has been focused on the application MAX phases as TBC materials[90,91]. The calculated values of the parameters related to the TBC applications for $Zr_2SeC$ are: $\Theta_D$ ~ 699.9 K, $TEC$ ~ $2.81\times10^{-6}K^{-1}$, $K_{min}$ – 1.3 W/mK, and $T_m$ ~ 1571 K. The values for a promising TBC material, $Y_4Al_2O_9$ are: $\Theta_D$ ~ 564 K, TEC ~ $7.51\times10^{-6}K^{-1}$, $K_{min}$ – 1.13 W/mK, and $T_m$ ~ 2020 K[92,93]. A comparison of these values indicates that the $Zr_2SeC$ MAX phase can be used as a TBC material for high-temperature applications.

*3.6 Optical properties*

The optical properties of the $Zr_2SeC$ compound are calculated and presented in Fig. 12 for [100] and [001] polarization directions up to 25 eV incident phonon energy. These two polarization directions represent the related electric field which is in perpendicular and parallel directions to the *c*-axis of unit cell structure. As confirmed that the titled compound is metallic; thus, a correction (intra-band)is needed to the low energy region of the spectrum. This correction is done by setting up a plasma frequency of3 eV and damping of 0.5 eV. Besides, Gaussian smearing of 0.5 eV was used with the intention that the k-points be more effective on the Fermi level. The optical constants of $Zr_2SeC$ are compared with those of $M_2SC$ (M = Zr, Hf, Nb) for [100] direction only.

Fig. 12 shows the (a) real and (b) imaginary part of the complex dielectric function of $Zr_2SeC$, which is very important and used to explain the material's response to the electric field [47]. As seen, the high negative static value of the real part and high positive static value of the imaginary part of the dielectric function indicates the metallic nature of $Zr_2SeC$; is consistent with the band structure result. The $\varepsilon_1(\omega)$ and $\varepsilon_2(\omega)$ of$Zr_2SeC$ are too much similar to those of $M_2SC$ (M = Zr, Hf, Nb). Strong anisotropy is observed in the low energy part, which tends to be isotropic in the high-energy region. The peaks in the $\varepsilon_2(\omega)$ are the results of electrons excitation, such as intra-band transitions (within M-$d$ states) and inter-band transitions due to absorption of photon incident on it. The first two peaks for [001] direction of $Zr_2SeC$ are due to electron transition within Zr-$d$ states. The third peak for [001] and the only peak for [100] direction assumed to be due to the inter-band transition of electrons. Approaching of zero by $\varepsilon_1(\omega)$ from below [at 14.91 and 15. 25 eV for [100] and [001] directions, respectively] and $\varepsilon_2(\omega)$ from above [at 19 eV for both directions] is another indication for the metallic nature of $Zr_2SeC$. The energy at which the $\varepsilon_1(\omega)$ touches zero from below are 15.42, 16.5, 16.44 eV for $M_2SC$ (M = Zr, Hf, Nb), respectively. Fig. 12(c) shows the refractive index $n(\omega)$ of $Zr_2SeC$. The value of static n(0) is 19 and 6 for [100] and [001]directions, respectively. As the value of $n(\omega)$ is an indication for the light velocity propagating the materials, thus, lower value of $n(\omega)$ is good for optical devices such as waveguides. The value of $n(\omega)$ is 6 for [001] direction implies the use of $Zr_2SeC$ in parallel to the field direction (parallel to the c-axis) is better than that of [100] direction (perpendicular to the c-axis). However, the velocity is reduced by the interaction with electrons during propagation through the $Zr_2SeC$, as confirmed from values greater than the unit value. Fig. 12(d) shows the imaginary part of the refractive index, known as extinction coefficient, $k(\omega)$, also used to measure the absorption capability of solids. The spectra of $k(\omega)$ follow the trend of $\varepsilon_2(\omega)$ for each of the MAX phases presented here. The static value of $k(\omega)$ is 2.8 and 0.23 for [100] and [001] directions, respectively. The $k(\omega)$ decreases to zero at ~ 16 eV, indicating that the intrinsic oscillation frequency of $Zr_2SeC$ is 16.0 eV. The higher value of $k(\omega)$ than $n(\omega)$ indicates the energy range wherein light cannot propagate, and $Zr_2SeC$ exhibit the behavior belongs to the reflective metal [94,95].

Fig. 12(e) shows the absorption coefficient (α) of $Zr_2SeC$ [100] and [001] for two polarization directions along with that of $M_2SC$ (M = Zr, Hf, Nb) for [100] direction only. The metallic behavior of $Zr_2SeC$ and $M_2SC$ (M = Zr, Hf, Nb) is reflected from the starting of the spectra, started at zero photon energy owing to no existence of the bandgap. For [100] directions, α increases with the rise in photon energy, exhibits a peak in the range of 1.5 – 1.9 eV, and then decreases slightly 3.3 to 3.6 eV, finally reaches maximum values at 7 to 9 eV. The energy range wherein α attains a value of greater than $10^5$ cm$^{-1}$ is the zone of strong absorption. The strong absorption zone of $Zr_2SeC$ is 3.5 to 14 eV and 2.8 to 14.4 eV for [100] and [001], respectively, indicating its application in the optoelectronic application within this energy range. The absorption coefficient of Zr$_2$SeC is similar to that of M$_2$SC (M = Zr, Hf, Nb) for [100] direction. The photoconductivity (σ) of Zr$_2$SeC is shown in Fig. 12(f), in which the curve is noted to be started in association with the photon incident, which implies the zero bandgap of the titled compound. The σ is found to be similar to that of $k(\omega)$ in peaks position as σ is the results of photon absorption that is exhibited by $k(\omega)$. Strong anisotropy is also observed in the low energy region that tends to be lowered in the high region.

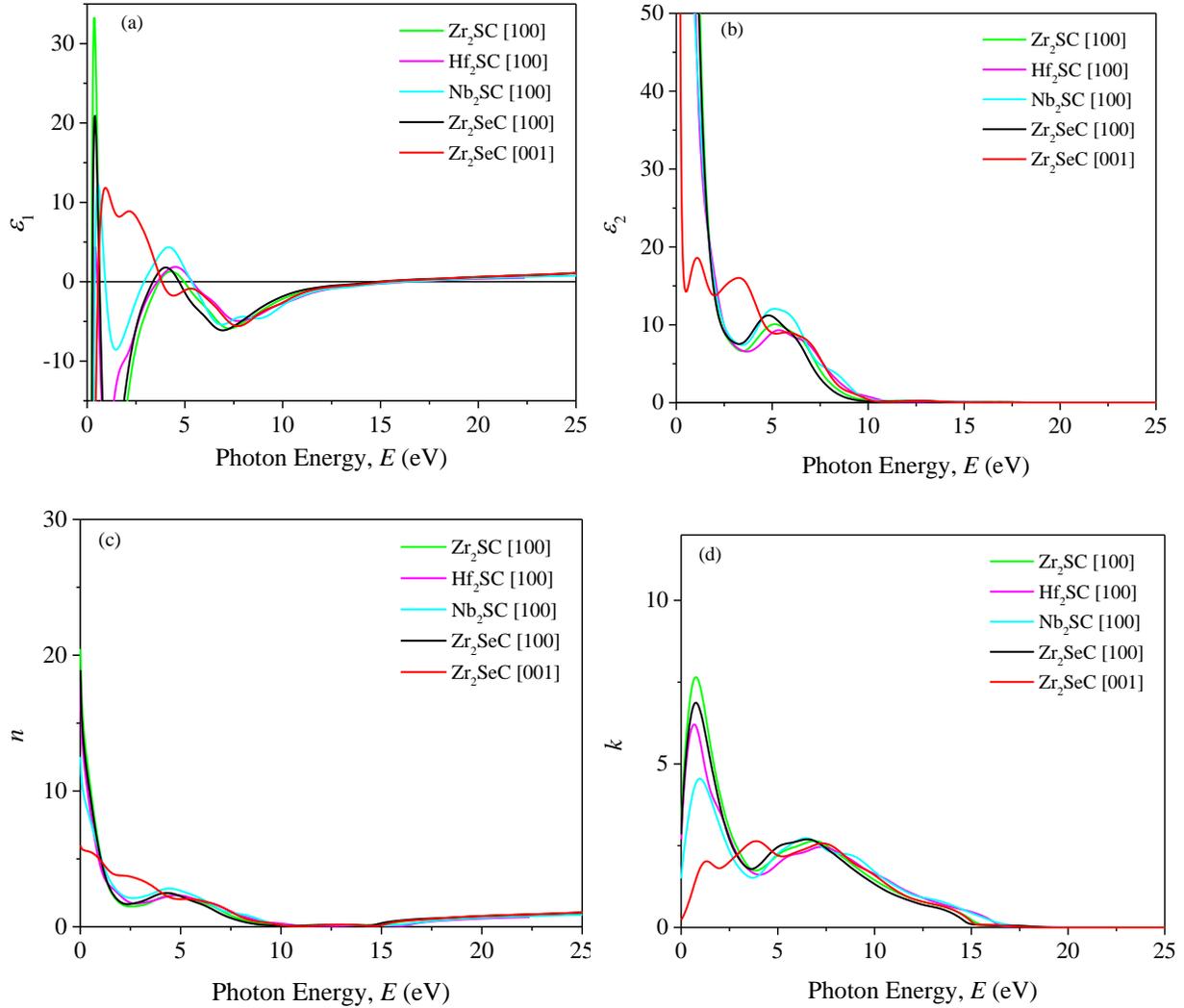

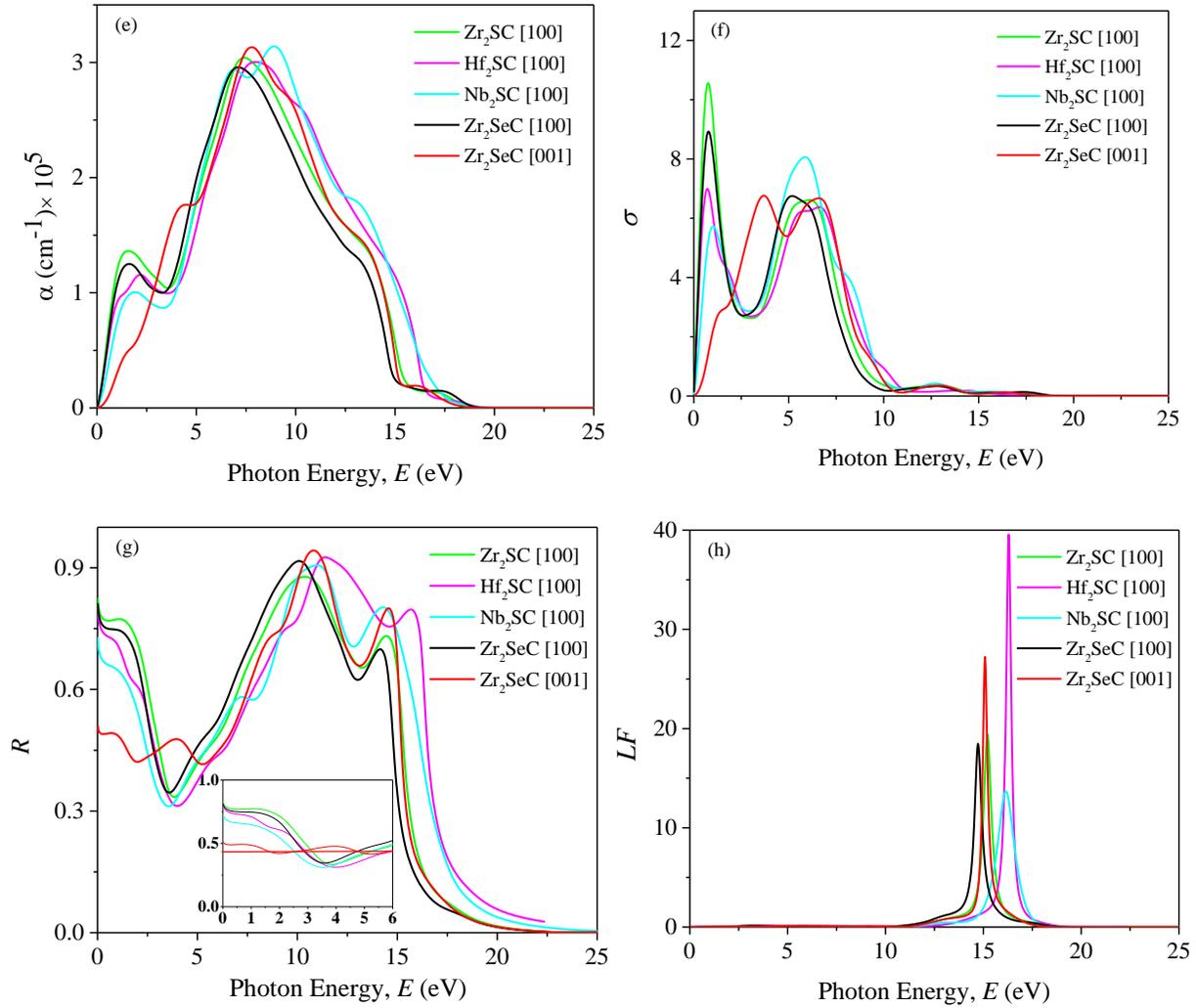

Fig. 12 (a) real part $\varepsilon_1$ and (b) imaginary part $\varepsilon_2$ of dielectric function $\varepsilon(\omega)$, (c) refractive index $n$, (d) extinction coefficient $k$, (e) absorption $\alpha$, (f) photoconductivity $\sigma$, (g) reflectivity $R$, and (h) loss function $LF$ of $Zr_2SeC$ together with those of $M_2SC$ (M = Zr, Hf, Nb) MAX phases as a function of photon energy.

Fig. 12(g) demonstrates the reflectivity ($R$) $Zr_2SeC$ that is similar to that of $M_2SC$ (M = Zr, Hf, Nb) for [100] direction. The static values of $R$ are 0.81 and 0.5 for [100] and [001] directions, respectively. The inset of Fig. 12(e) shows the $R$ up to 6 eV in which a horizontal redline is drawn at 0.44 (44%) of R to demonstrate the suitability for use as a cover material to diminish solar heating. It was suggested by Li et al.[96,97] that the MAX phase with reflectivity greater than 44% is suitable for use cover materials for the same mentioned before. As seen, $R$ decreases to lower than 0.44 at 1.7 and 2.8 eV for [001] and [100] directions, respectively. However, $R$ of Zr$_2$SeC decreases to lower than 0.44 at slightly lower energy than that of M$_2$SC (M = Zr, Hf), but the lowest value (34 %) is higher than that of M$_2$SC (M = Zr, Hf, Nb). $R$ increases to a maximum of 0.91at 10 eV, followed by another peak of 0.7 at 14.1 eV and finally declined sharply towards zero, and the titled compound becomes transparent for the incident light. The Zr$_2$SeC shows reflective behavior in the energy range of 7.5 to 14 eV in which the refractive index is very low, and the reflectivity is greater than 70%. A similar reflectivity is also shown by M$_2$SC (M = Zr, Hf, Nb) MAX phases. Fig12 (h) shows the loss function of Zr$_2$SeC that demonstrates how fast electrons losing their energy when traversing the Zr$_2$SeC compound. A peak corresponding to the plasma frequency ($\omega_p$) is observed at the energy where $\varepsilon_1(\omega)$ and $\varepsilon_2(\omega)$ approaches zero from below and above respectively, and reflectivity shows trailing edges. The $\omega_p$ of Zr$_2$SeC are 14.7 and 15.1 eV for [001] and [100] directions, respectively. The $\omega_p$ of M$_2$SC (M = Zr, Hf, Nb) MAX phases are slightly higher than that of Zr$_2$SeC for [100] direction. This plasma frequency defined a critical value at which the material changes its behavior from metallic to transparent dielectric.

## 4. Conclusions

A comparative study of the structural, electronic, mechanical, thermal, and optical properties of synthesized MAX phase Zr$_2$SeC with prior known MAX phase Zr$_2$SC has been performed by density functional theory for the first time. The obtained lattice constants, volume, and atomic positions are noticed to be agreed with experimental and theoretical results. The metallic behavior has been confirmed from the band structure and density of states (DOS). The less energy dispersion along the *c*-direction confirmed the anisotropic behavior of electrical conductivity, and partial DOS revealed the dominant role of Zr-*d* states to the electrical conductivity. Shifting of the peak in the DOS owing to the replacement of Se by S and difference in charge density mapping (CDM) revealed that the strength of bonds present in Zr$_2$SeC is lower than the bonds in Zr$_2$SC. The mechanical stability of the synthesized phase is further confirmed by the stiffness constants. The elastic constants, elastic moduli, and hardness parameters of Zr$_2$SeC are found lower than those of Zr$_2$SC. The lowering of the parameters is explained based on bond populations, and bond lengths present within them in an association with the DOS and CDM results. Both the directional projections of the elastic moduli and anisotropy indices confirm the anisotropic behavior of Zr$_2$SeC and Zr$_2$SC; Zr$_2$SeC is more anisotropic than Zr$_2$SC.

The temperature and pressure-dependent properties exhibit the expected variation for the same. The value of $\Theta_D$, $K_{min}$, $T_m$, and $TEC$ of $Zr_2SeC$ are lower than those of $Zr_2SC$, $C_v$ of $Zr_2SeC$ reach classical DP limit at a lower temperature than that of $Zr_2SC$; is consistent with the parameters used to characterize mechanical properties. A comparison of the values of $\Theta_D$, $K_{min}$, $T_m$, and $TEC$ for $Zr_2SeC$ with those of a promising TBC material ($Y_4Al_2O_9$) [93] reveals its possibility to be used as TBC material. The real and imaginary parts of the dielectric constant and the absorption and photoconductivity curves agree with the band structure results and confirm the metallic nature of $Zr_2SeC$. The possible relevance of $Zr_2SeC$ for use in optoelectronic applications and as shielding material to diminish solar heating has been confirmed from the analysis of the studied optical properties. Like electrical conductivity and mechanical properties, the optical properties of $Zr_2SeC$ also exhibit anisotropic behavior.